\documentstyle[12pt,prl,aps]{revtex}

\def\beq{\begin{equation}}
\def\eeq{\end{equation}}

\newcommand{\comb}[2]{\Bigl({\textstyle{#1\atop#2}}\Bigr)}
\newcommand{\fract}[2]{\textstyle{\frac{#1}{#2}}}
\newcommand{\ket}[1]{| #1 \rangle}
\newcommand{\tket}[1]{\left|#1\right\rangle}
\let\hat\widehat

\draft
\tighten

\begin{document}

\title{How many functions can be distinguished\\
with \protect$\bbox{k}$ quantum queries?\cite{TitleThanks}}

\author{Edward Farhi and Jeffrey
Goldstone\cite{FarhiGoldstoneAuth}}
\address{Center for Theoretical Physics\\ 
Massachusetts Institute of Technology\\
Cambridge, MA  02139}

\author{Sam Gutmann\cite{GutmannAuth}} 

\address{Department of Mathematics\\ 
Northeastern University\\ 
Boston, MA 02115}

\author{Michael Sipser\cite{SipserAuth}}

\address{Department of Mathematics\\ 
Massachusetts Institute of Technology\\
Cambridge, MA  02139 \\[2em]
{\rm MIT-CTP-2814~~quant-ph/9901012 \hfil \qquad \qquad January 1999}\\[3em]}

\maketitle

\begin{abstract} Suppose an oracle is known to hold one of a
given set of $D$ two-valued functions. To successfully
identify which function the oracle holds with $k$ classical
queries, it must be the case that $D$ is at most $2^k$. In
this paper we derive a bound for how many
functions can be distinguished with $k$ quantum queries.
\end{abstract}

\section{Introduction} Quantum computers can solve certain
oracular problems with fewer queries of the oracle than are
required classically. For example, Grover's
algorithm\cite{bib:1} for unstructured search can be viewed
as distinguishing between the $N$ functions
\beq G_j(x) =\left\{ \begin{array}{rl}
                        -1 &\quad \mbox{ for $x=j$}\\
                         1&\quad \mbox{ for $x\neq j$}
                       \end{array} \right.
\label{eq:1}
\eeq where both $x$ and $j$ run from 1 to $N$. Identifying
which of these $N$ functions the oracle holds
requires of order~$N$ queries classically, whereas quantum mechanically
this can be done with of order~$\sqrt N$ quantum queries.

The $N$ functions in (\ref{eq:1}) are a subset of the $2^N$
functions 
\beq F: \{1,2,\ldots,N\} \rightarrow \{-1,1\}\ . 
\label{eq:2}
\eeq
\emph{All} $2^N$ functions of this form can be distinguished
with $N$ queries so the $N$ functions in (\ref{eq:1}) are
particularly hard to distinguish classically. 
No more than $2^k$ functions can be distinguished
with only $k$ classical queries,
since each query has only two possible
results. Note that this classical ``information'' bound of
$2^k$ does not depend on $N$, the size of the domain of the
functions. 

Quantum mechanically the $2^k$ information bound does not
hold\cite{bib:2}. In this paper we derive an upper bound for
the number of functions that can be
distinguished with $k$ quantum queries. If there is a set of
$D$ functions  of the form (\ref{eq:2}) that can be
distinguished with
$k$ quantum queries, we show that
\beq D \le 1 + \comb N1 + \comb N2 + \cdots + \comb Nk \ .
\label{eq:3}
\eeq If the probability of successfully identifying which
function the oracle holds is only required to be $p$ for
each of the $D$ functions, then
\beq D \le \fract1p \left[ 1 + \comb N1 + \comb N2 + \cdots
+ \comb Nk\right] \ .
\label{eq:4}
\eeq

We also give two examples of sets of $D$ functions
(and values of $k$ and $p$) where (\ref{eq:3}) and
(\ref{eq:4}) are equalities. In these cases the quantum
algorithms succeed with fewer queries than the best
corresponding classical algorithms. One of these examples shows
that van Dam's algorithm\cite{bib:4} distinguishing all $2^N$
functions with high probability after $N/2+O(\sqrt{N})$ queries
is best possible, answering a question posed in his paper.
We also give an example
showing that the bound (\ref{eq:3}) is not always tight. 

An interesting consequence of (\ref{eq:3}) is a lower bound on
the number of quantum queries needed to sort $n$ items in the
comparison model. Here, we have $D=n!$ functions, corresponding
to the $n!$ possible orderings, to be distinguished. The
domain of these functions is the set of $N=\comb n2$ pairs
of items.  If $k = (1-\epsilon)n$, the bound (\ref{eq:3}) 
is violated for $\epsilon>0$ and $n$ large, as is easily checked.
Hence, for any $\epsilon>0$ and $n$ sufficiently large, $n$ items
cannot be sorted with $(1-\epsilon)n$ quantum queries.

\section{Main result} Given an oracle associated with any
function $F$ of the form (\ref{eq:2}), a quantum query is an
application of the unitary operator, $\hat F$, defined by 
\beq
\hat F \tket{x,q,w} = \tket{x, q\cdot F(x), w}
\label{eq:5}
\eeq where $x$ runs from 1 to $N$, $q=\pm1$, and $w$ indexes
the work space. A quantum algorithm that makes $k$ queries
starts with an initial state $\ket s$ and alternately
applies $\hat F$ and $F$-independent unitary operators,
$V_i$, producing
\beq
\ket{\psi_F} = V_k \hat FV_{k-1}\cdots V_1\hat F \ket s \ .
\label{eq:6}
\eeq

Suppose that the oracle holds one of the $D$ functions $F_1,
F_2,\ldots,F_D$, all of the form  (\ref{eq:2}). If the
oracle holds $F_j$, then the final state of the algorithm is
$|{\psi_{F_j}}\rangle$, but we do not (yet) know what $j$
is. To identify $j$ we divide the Hilbert space into $D$
orthogonal subspaces with corresponding projectors $P_1,
P_2, \ldots, P_D$. We then make simultaneous measurements
corresponding to this commuting set of projectors. One and
only one of these measurements yields a~$1$. If the $1$ is
associated with $P_\ell$ we announce that the oracle holds 
$F_\ell$. 

Following\cite{bib:3}, if the oracle holds $F$, so that the
state before the measurement was $\ket{\psi_F}$ given by
(\ref{eq:6}), we know that for each~$\ell$, $\left\| P_\ell
\ket{\psi_F}\right\|^2$ is a $2k$-degree polynomial in the
values $F(1), F(2),\ldots,F(N)$. More precisely, 
\beq
\Bigl\| P_\ell \ket{\psi_F}\Bigr\|^2 =
\sum_{r=1}^{m_\ell}\Bigl | Q_{\ell r}\left( F(1),
\ldots, F(N) \right)\Bigr|^2
\label{eq:7}
\eeq
where each $ Q_{\ell r}$ is a $k$-th degree multilinear polynomial
and $m_\ell$ is the dimension of the  $\ell$-th subspace. 
Note that formula  (\ref{eq:7}) holds for \emph{any} $F$
whether or not $F=F_j$ for some~$j$.  The algorithm
succeeds, with probability at least~$p$, if for each
$j=1,\ldots,D$, we have
\beq
\Bigl\| P_j |\psi_{F_j}\rangle\Bigr\|^2 = \sum_{r=1}^{m_j}
\Bigl| Q_{j r}\left( F_j(1),
\ldots, F_j(N) \right)\Bigr|^2 \ge p\ . 
\label{eq:8}
\eeq

We now prove the following lemma: Let $F_0$ be any one of
the functions of form (\ref{eq:2}). If $Q$ is a polynomial
of degree at most~$k$ such that
\beq
\Bigl| Q\left( F_0(1),
\ldots, F_0(N) \right)\Bigr|^2 =1
\label{eq:9}
\eeq then 
\beq
 \sum_F \Bigl| Q\left( F(1),
\ldots, F(N) \right)\Bigr|^2 \ge \frac{2^N}{1+\comb N1 +
\cdots + \comb Nk}
\label{eq:10}
\eeq where the sum is over all $2^N$ functions of the form
(\ref{eq:2}). Proof: Without loss of generality we can take
$ F_0(1) = F_0(2) = \cdots = 
 F_0(N)= 1$. Now
\beq
 Q\left( F(1), \ldots, F(N) \right) =  a_0 + \sum_x a_x F(x)
+ \sum_{x<y} a_{xy} F(x) F(y) + \cdots
\label{eq:11}
\eeq where the last term has $k$ factors of $F$ and the
coefficients are complex numbers. Note that
\beq
\sum_F F(x_1)F(x_2)\cdots F(x_g) F(y_1)\cdots F(y_h) = 0
\label{eq:12}
\eeq as long as the sets $\{ x_1, \ldots, x_g\}$ and $\{y_1,
\ldots, y_h\}$ are not equal and $x_1,\ldots,x_g$ are
distinct, as are $y_1, \ldots, y_h$. This means that
\beq
\sum_F \Bigl| Q\left( F(1), \ldots, F(N) \right)\Bigr|^2 = 
2^N \Big( |a_0|^2 + \sum_x |a_x|^2 + \sum_{x<y} |a_{xy}|^2 + \cdots\ \Big)
.
\label{eq:13}
\eeq Now (\ref{eq:9}) with $F_0(x)\equiv1$ means 
\beq
\Bigl| a_0 + \sum_x a_x + \sum_{x<y} a_{xy} + \cdots
\Bigr|^2 =1 \ .
\label{eq:14}
\eeq Because of the constraint (\ref{eq:14}), the minimum
value of (\ref{eq:13}) is achieved when all the coefficients
are equal. Since there are $1+\comb N1 + \cdots +
\comb Nk$ coefficients, the inequality (\ref{eq:10}) is established.

Suppose we are given an algorithm that meets condition
(\ref{eq:8}) for
$j=1,\ldots,D$. Then by the above lemma,
\beq
 \sum_{r=1}^{m_j} \sum_F \Bigl| Q_{j r}\left( F(1),
\ldots, F(N) \right)\Bigr|^2 \ge \frac{2^N p}{1+\comb N1 +
\cdots + \comb Nk} \ . 
\label{eq:15}
\eeq Summing over $j$ using (\ref{eq:7}) yields
\beq
 \sum_j \sum_F \Bigl\| P_j\ket{\psi_F}\Bigr\|^2 \ge \frac{D
2^N p}{1+\comb N1 +
\cdots + \comb Nk} \ . 
\label{eq:16}
\eeq For each $F$, the sum on $j$ gives 1 since $\Bigl\|
\tket{\psi_F} \Bigr\| =1$. Therefore the lefthand side of
(\ref{eq:16}) is $2^N$ and (\ref{eq:4}) follows. 

\section{Examples}

\leavevmode\llap{0.\ }  If $k=N$, all $2^N$ functions can be
distinguished classically and therefore quantum
mechanically. In this case (\ref{eq:3}) becomes
\beq
2^N = D \le 1 + \comb N1 + \comb N2 + \cdots + \comb NN =
2^N\ .
\label{eq:17}
\eeq 
\bigskip

\leavevmode\llap{1.\ }  For $k=1$, if $N=2^n-1$ there are
$N+1$ functions that can be distinguished\cite{bib:2} so the
bound (\ref{eq:3}) is best possible. The functions can be
written as
\beq f_a(x) = (-1)^{a\cdot x}
\label{eq:18}
\eeq with $x\in \{1,\ldots,N\}$ and $a\in \{0, 1,\ldots,N\}$
and $a\cdot x = \sum_i a_i x_i$ where $a_1\cdots a_n$ and
$x_1\cdots x_n$ are the binary representations of
$a$ and $x$. To see how these can be distinguished we work
in a Hilbert space with basis $\{\tket{x,q}\}$, $x=1,\ldots,N$
and $q=\pm 1$, where a quantum query is defined as in (\ref{eq:5})
and the work bits have been suppressed.  We define
\beq \label{eq:19}
\tket{x}=\frac{1}{\sqrt{2}}\{\tket{x,+1}-\tket{x,-1}\} \quad
  x=1,\ldots,N
\eeq
and
\beq \label{eq:20}
\tket{0}=\frac{1}{\sqrt{2}}\{\tket{1,+1}+\tket{1,-1}\} ,
\eeq
so $\{\tket{x}\}$, $x=0,1,\ldots,N$, is an orthonormal set.
Now by (\ref{eq:5}), if we define $F(0)$ to be $+1$, we have
\beq \label{eq:21}
\hat F\tket{x} = F(x)\tket{x} \quad x=0,1,\ldots,N
\eeq
and in particular,
\beq \label{eq:22}
\hat f_a\tket{x} = (-1)^{a\cdot x}\tket{x} \quad x=0,1,\ldots,N
\eeq
Now let
\beq
 \tket{s} =  \frac1{\sqrt{N+1}} \sum_{x=0}^N\tket{x} 
\label{eq:23}
\eeq and observe that the $N+1$ states $\hat f_a\tket{s}$
are orthogonal for
$a=0,1,\ldots,N$. 
\bigskip

\leavevmode\llap{2.\ } In\cite{bib:4} an algorithm is
presented that distinguishes all $2^N$ functions in $k$
calls with probability $\left( 1 + \comb N1 + \cdots + \comb
Nk \right) \bigm/ 2^N$.  With this value of $p$, and $D=2^N$, the
bound (\ref{eq:4}) becomes an equality. Furthermore, (\ref{eq:4})
shows that this algorithm is best possible. 
\bigskip

\leavevmode\llap{3.\ } Nowhere in this paper have we 
exploited the fact  that for an algorithm that succeeds with
probability~1, it must be the case that  $\left\| P_\ell
\ket{\psi_{F_j}}\right\|=0$ for
$\ell\neq j$. With this additional constraint it can be
shown that for $N=3$,  no set of $7 = 1
+ \comb31 + \comb32$ functions can be distinguished with $2$
quantum queries. Thus for $N=3$ and $k=2$ the bound (\ref{eq:3})
is not tight.

\end{document}